\documentclass[slac_one]{revtex4}
\usepackage{graphicx}
\usepackage{fancyhdr}
\renewcommand{\headrulewidth}{0pt}
\renewcommand{\footrulewidth}{0pt}

\begin{document}

\title{Early B-physics at ATLAS} 

\author{S. Oda on behalf of the ATLAS Collaboration}
\affiliation{International Center for Elementary Particle Physics, The University of Tokyo, 7-3-1 Hongo, Bunkyo-ku, Tokyo 113-0033, Japan}

\begin{abstract}
The $B$-physics program at the ATLAS experiment, which covers the mid-rapidity region, complements that at the dedicated LHCb experiment, which covers the forward rapidity region. 
At the early stage of the LHC operation, the program concentrated on understanding of detector performance and measurements of quarkonia and $D$ mesons. 
This article presents recent results of the $B$-physics program at ATLAS. 
\end{abstract}

\maketitle

\thispagestyle{fancy}

\section{ATLAS B-PHYSICS PROGRAM} 
The $B$-physics program at the ATLAS experiment at the LHC evolves with the integrated luminosity and is summarized in~\cite{ATLAS-CSC-NOTE}. 
At the early stage of the program with the integrated luminosity of less than $10$~pb$^{-1}$, ATLAS concentrates on understanding detector performance using well understood $c$- and $b$-processes and measurements of production cross sections for $D$-hadrons, $B$-hadrons, $J/\psi$ and $\Upsilon$ to test QCD predictions for proton-proton ($pp$) collisions at $\sqrt{s}=7$~TeV. 
ATLAS will collect $1$~fb$^{-1}$ data by the end of 2011. 
With the high statistics, contribution to world averages on the properties of $B$-hadrons ($B$, $B_s$, $B_c$, $\Lambda_b$) will start and quarkonium polarization will be measured. 
With more statistics from 2013, rare decays (such as $B_s\rightarrow\mu^+\mu^-$) and Beyond Standard Model CP-violation in weak decays of $B$ mesons will be searched, and $\Lambda_b$ polarization will be measured. 
ATLAS recorded $1.544$~pb$^{-1}$ data as of August 20, 2010 and analyses use different subsets of the recorded data sample. 

\section{ATLAS DETECTOR AND TRIGGER}
The ATLAS detector is a general purpose detector at the LHC and covers the rapidity ($y$) from $-2.4$ to $+2.4$~\cite{ATLAS-DET}. 
It complements the $B$-physics program of the dedicated LHCb experiment covering $2<y<4.5$. 
The Inner Detector (ID) is immersed in a $2$~T solenoid magnet and consists of Silicon Pixels and Silicon Strips (SCT), which provide precise tracking and vertexing. 
The Muon Spectrometer (MS) consists of gas-based chambers and air-core toroid magnets. 
MS provides muon triggers and momentum measurements with $<10$\% resolution up to $1$~TeV. 

Muons are reconstructed either fully or partially. 
Fully reconstructed muons, which are called combined muons, have an ID track matched to a MS track and refitted through the detector to give the best measurement. 
Partially reconstructed muons, called tagged muons, are ID tracks matched to muon segments when extrapolated to MS. 
Tagged muons generally have low transverse momenta ($p_T$). 

The ATLAS trigger consists of three levels, Level 1 (L1), Level 2 (L2) and Event Filter (EF). 
L1 trigger is a hardware trigger based on MS, Calorimeters and Minimum Bias Trigger Scintillator (MBTS). 
L2 and EF triggers are software triggers and called High Level Trigger (HLT). 
The L2 trigger is used to confirm L1 trigger decision and EF trigger is to perform event selection using more complex algorithms. 
There are three kinds of HLT $B$-triggers, full scan trigger, single muon trigger and di-muon trigger. 
The di-muon mass limit is set at $13$~GeV to cover quarkonium decays ($J/\psi$, $\psi'$, $\Upsilon\rightarrow\mu^+\mu^-$), rare $B$-decays ($B_{s, d}\rightarrow\mu^+\mu^-$, $B\rightarrow X_{s}\mu^+\mu^-$). 
The trigger menu evolves with the instantaneous luminosity and well-understood triggers are used for physics analysis. 

\section{PERFORMANCE STUDY}
Reconstructed invariant mass distributions of $D$, $J/\psi$ and $\Upsilon$ mesons are used to study Inner Detector performance. 

\subsection{Using D mesons} 
The reconstruction of $D$ mesons is studied with L1 MBTS triggered data corresponding to the integrated luminosity of $1.4$~nb$^{-1}$~\cite{ATLAS-CONF-2010-034}. 
The reconstruction is based on ID tracks with pion or kaon mass hypothesis without particle identification. 

\begin{figure*}[t]
\centering
\includegraphics[width=74mm]{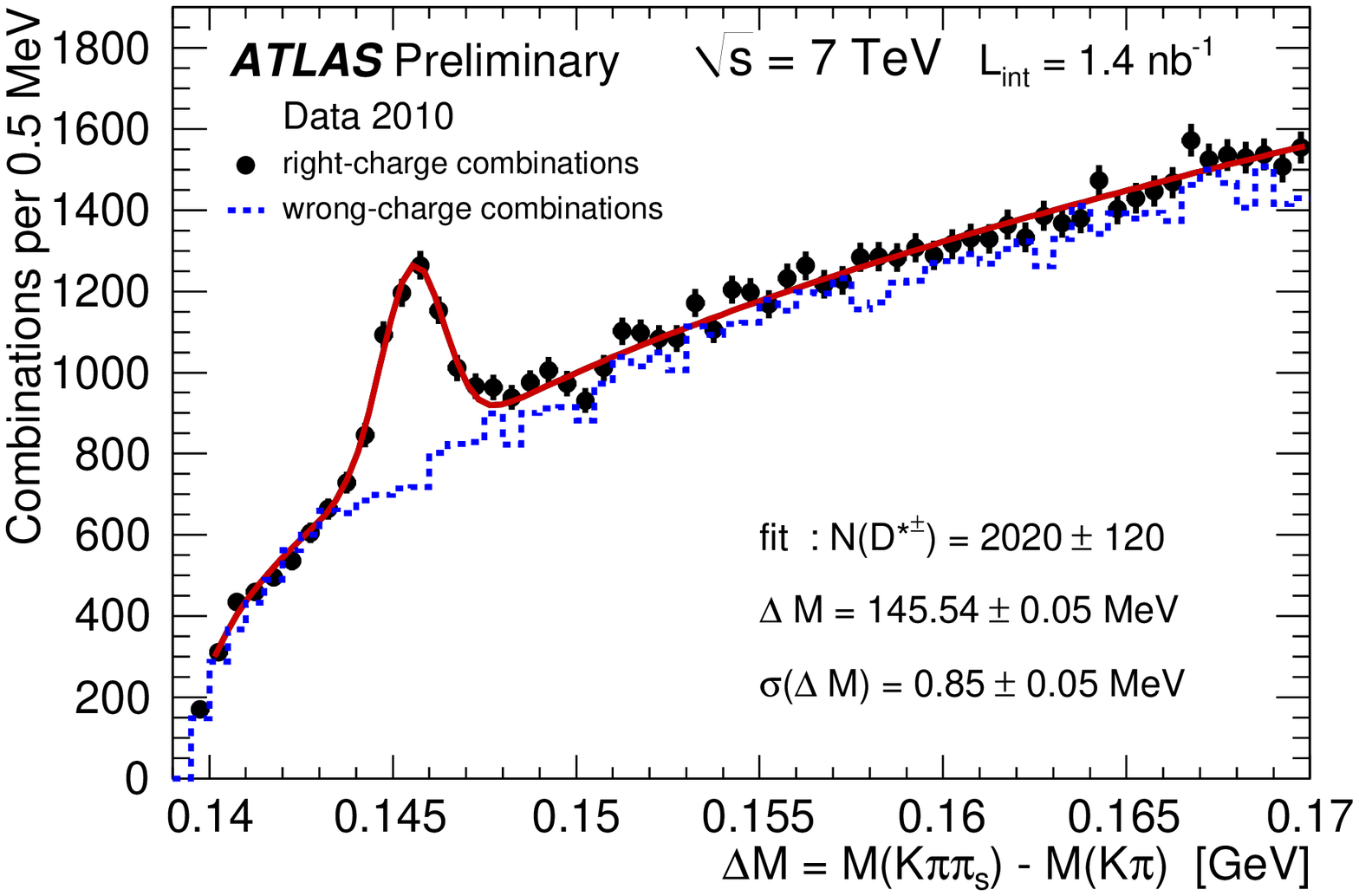}
\includegraphics[width=74mm]{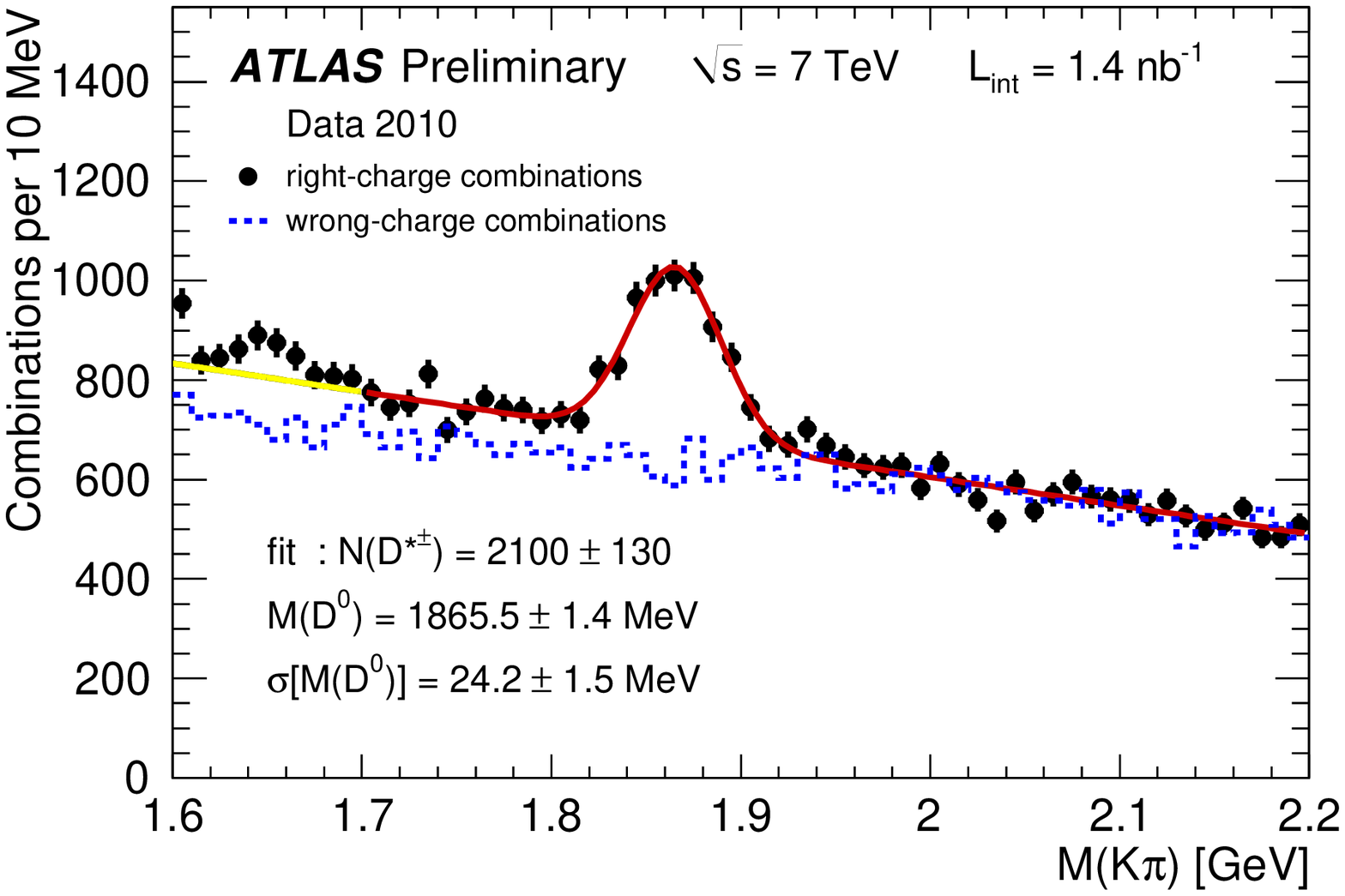}
\includegraphics[width=74mm]{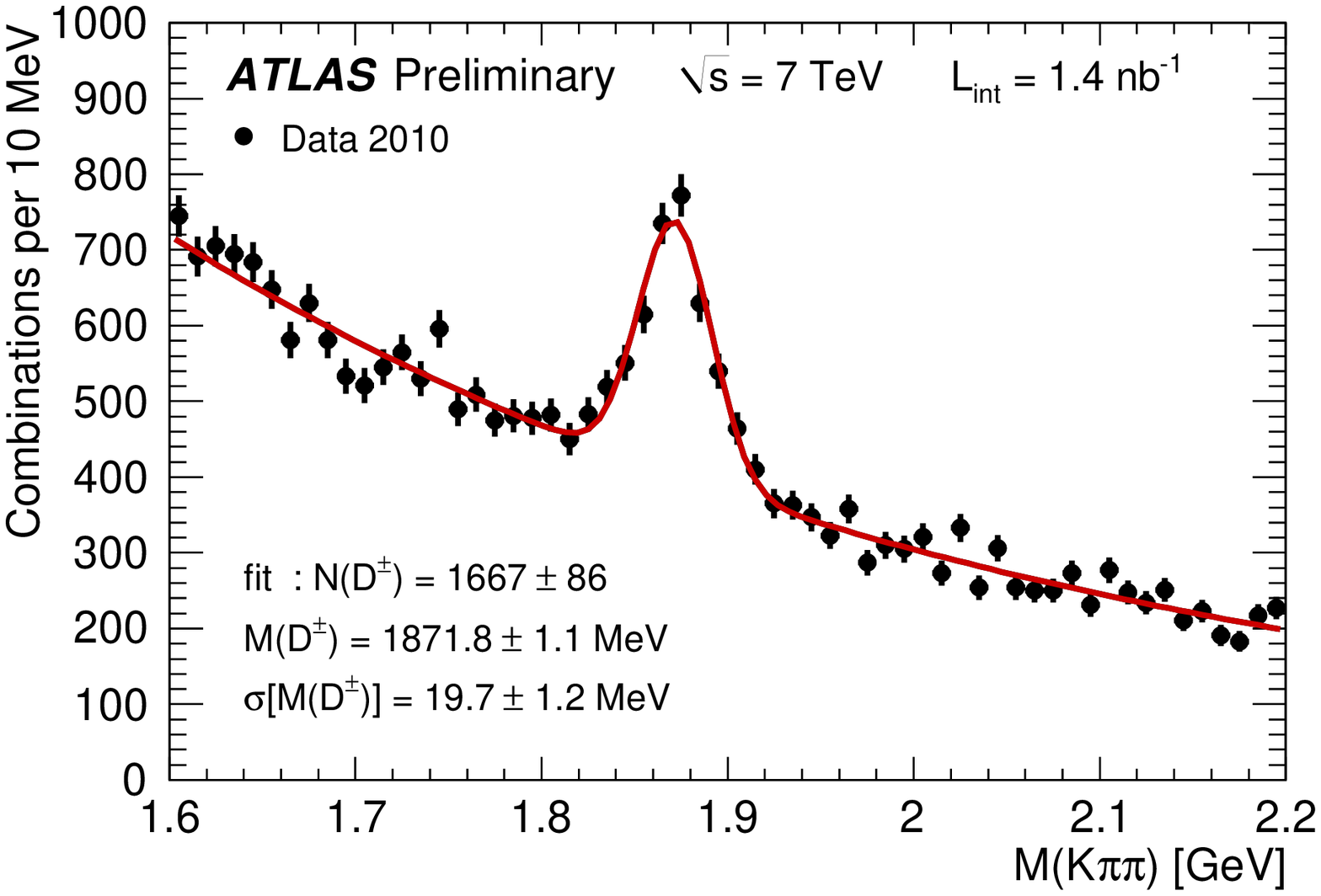}
\includegraphics[width=74mm]{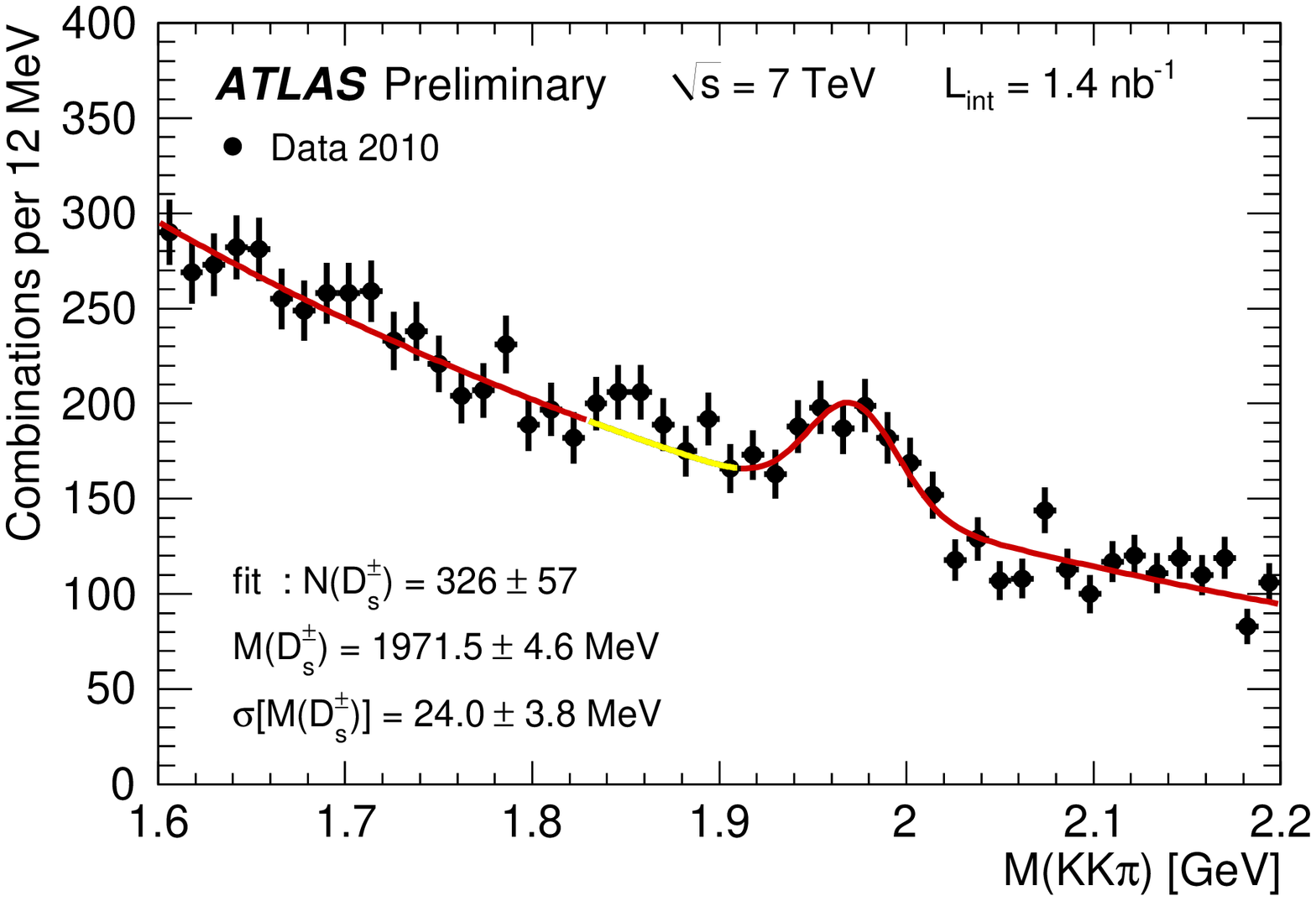}
\caption{The distribution of the mass difference, $\Delta M=M(K\pi\pi_s)-M(K\pi)$, (top left) and the distribution of $M(K\pi)$ (top right) for the $D^{*\pm}$ candidates. 
The distribution of $M(K\pi\pi)$ for the $D^\pm$ candidates (bottom left) and one of $M(KK\pi)$ for the $D^\pm_s$ candidates (bottom right). 
} \label{fig_01}
\end{figure*}

\subsubsection{$D^{*+}\rightarrow D^{0}\pi^{+}_s\rightarrow(K^{-}\pi^{+})\pi^{+}_s$ and the charge conjugate} 
Two opposite sign tracks with $p_T>1$~GeV are combined with kaon and pion masses. 
A third track with $p_T>0.25$~GeV is combined with pion mass and is called soft pion ($\pi_s$) due to the small mass difference between $D^{*+}$ and $D^0$. 
Candidates of $D^{*+}$ mesons are identified with the mass difference ($\Delta M=M(K\pi\pi_s)-M(K\pi)$). 
There are $2020 \pm 120$ $D^{*}$ candidates satisfying $144<\Delta M<147$~MeV (see the top left panel of Fig.~\ref{fig_01}). 
The mean mass difference is $145.54 \pm 0.05$~MeV (all the errors quoted in the mass fits in this article are statistical only) is close to the PDG value ($145.42 \pm 0.01$~MeV). 
With the $D^{*+}$ selection, $2100 \pm 130$ $D^{0}$ candidates are found in the mass region of $1.83<M(K\pi)<1.90$~GeV (see the top right panel of Fig.~\ref{fig_01}). 
The width of the signal is $24.2 \pm 1.5$~MeV in agreement with the Monte Carlo (MC) detector simulation expectation. 

\subsubsection{$D^{+}\rightarrow K^{-}\pi^{+}\pi^{+}$ and the charge conjugate} 
Two same sign tracks with $p_T>0.8$~GeV are combined with pion mass. 
An opposite sign track with $p_T>1$~GeV is combined with kaon mass. 
To suppress the contribution of $D^{*+}\rightarrow D^{0}\pi^{+}_s\rightarrow(K^{-}\pi^{+})\pi^{+}_s$, $\Delta M=M(K\pi\pi)-M(K\pi)>150$~MeV is required. 
To suppress the contribution of $D_s^+\rightarrow \phi\pi^+\rightarrow(K^+K^-)\pi^+$, $|M(K^+K^-)-M(\phi)_{PDG}|>8$~MeV is required. 
As shown in the bottom left panel of Fig.~\ref{fig_01}, the fitted yield is $1667 \pm 86$. 
The fitted mass value, $1871.8\pm1.1$~MeV, is in good agreement with the PDG value ($1869.6\pm0.2$~MeV). 
The width of the signal is $19.7 \pm 1.2$~MeV in agreement with the MC expectation. 

\subsubsection{$D_s^+\rightarrow \phi\pi^+\rightarrow(K^+K^-)\pi^+$ and the charge conjugate}
Two opposite sign tracks with $p_T>0.7$~GeV are combined with kaon mass. 
If $|M(KK)-M(\phi)_{PDG}|$ is less than $6$~MeV, the pair is considered as a good $\phi$ meson candidate. 
A third track with $p_{T}>0.8$~GeV is combined with pion mass. 
As shown in the bottom right panel of Fig.~\ref{fig_01}, the fitted $D_s$ yield is $326 \pm 57$ and the fitted mass value, $1971.5 \pm 4.6$~MeV, is consistent with the PDG value ($1968.5 \pm 0.3$~MeV). 
The region of $D^{+}\rightarrow K^+K^-\pi^+$ is excluded from the fit. 

\subsection{Using $J/\psi$ meson} \label{withJpsi}
\begin{figure*}[hb]
\centering
\includegraphics[width=58mm]{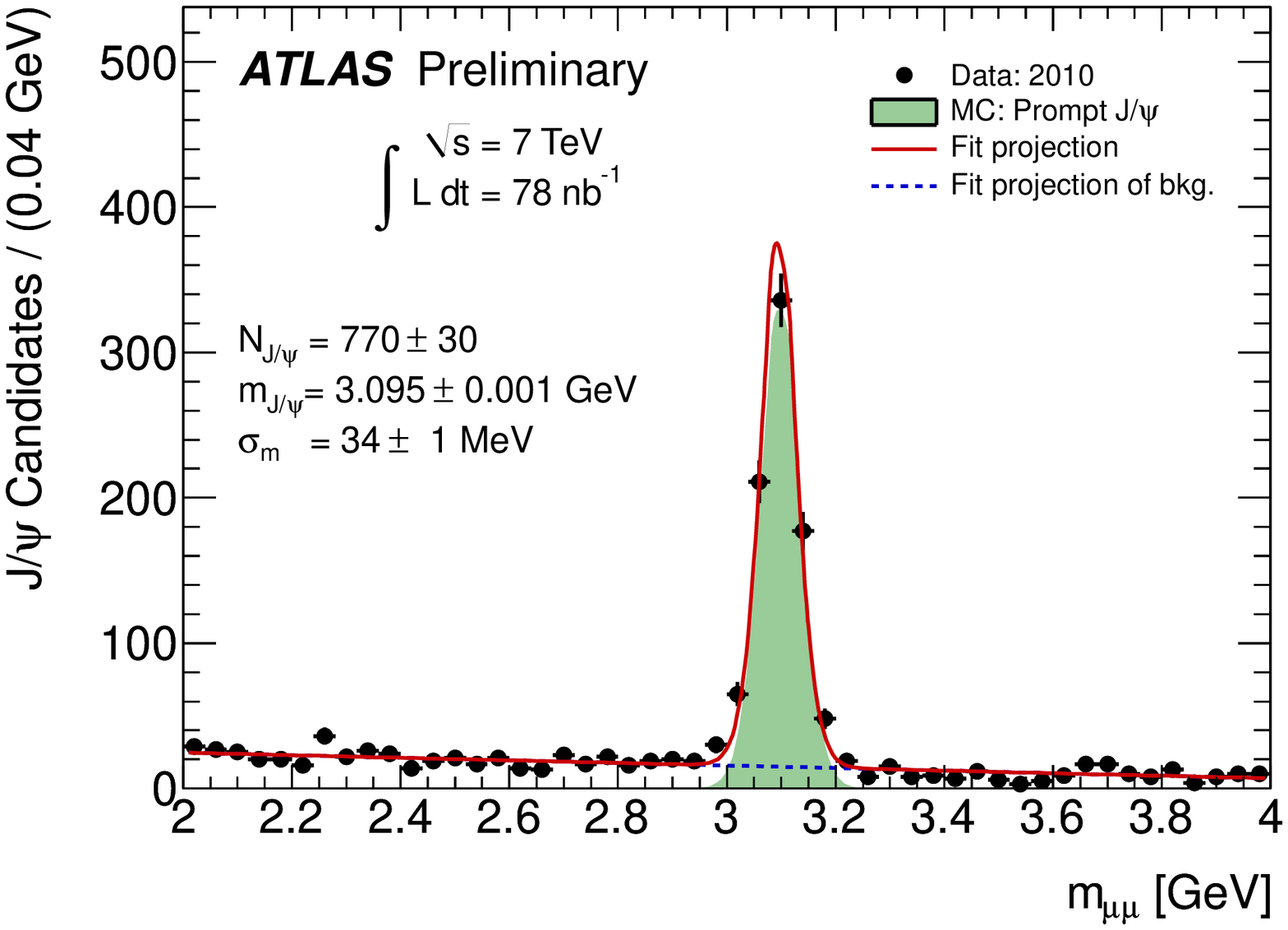}
\includegraphics[width=58mm]{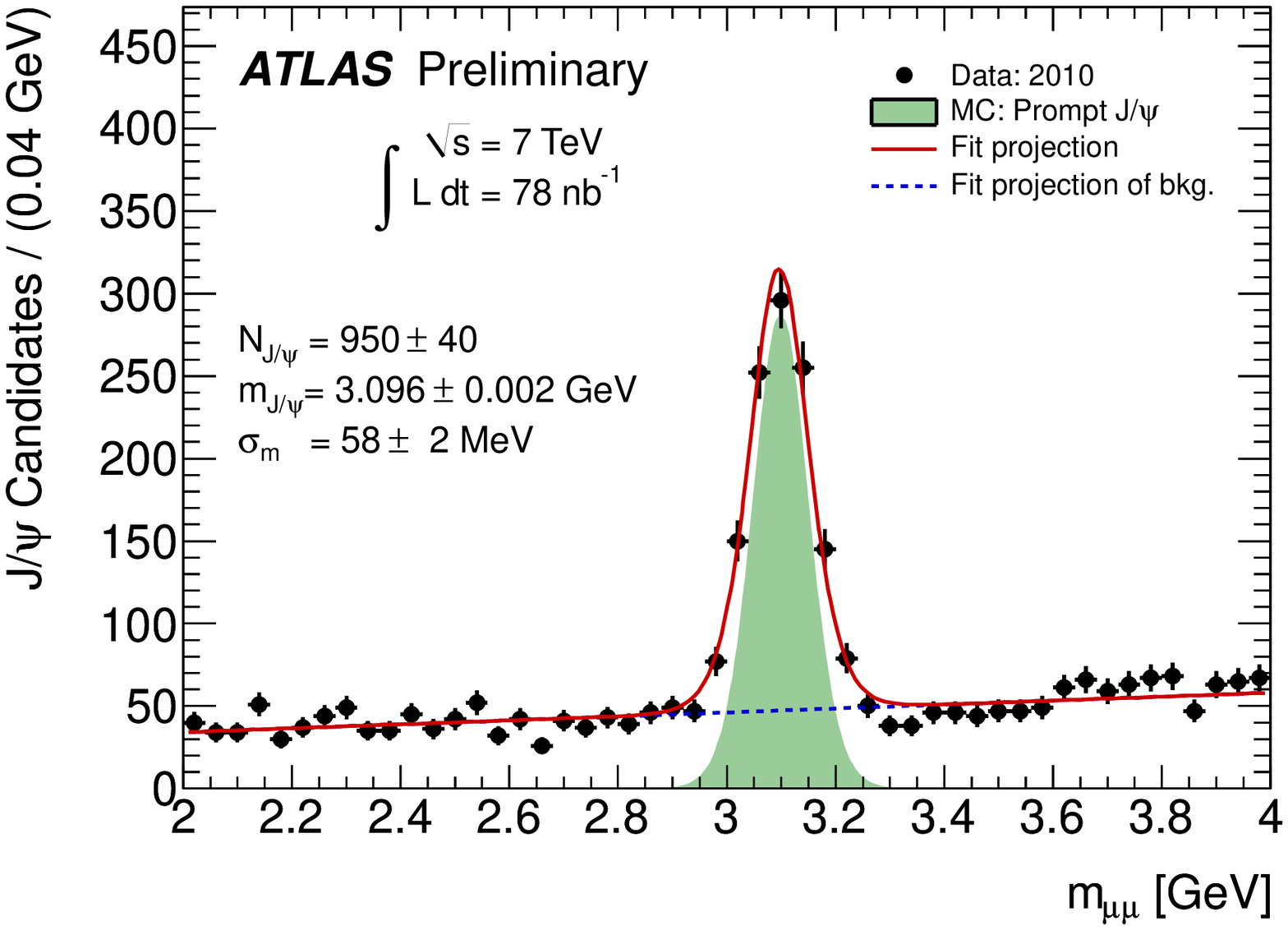}
\includegraphics[width=58mm]{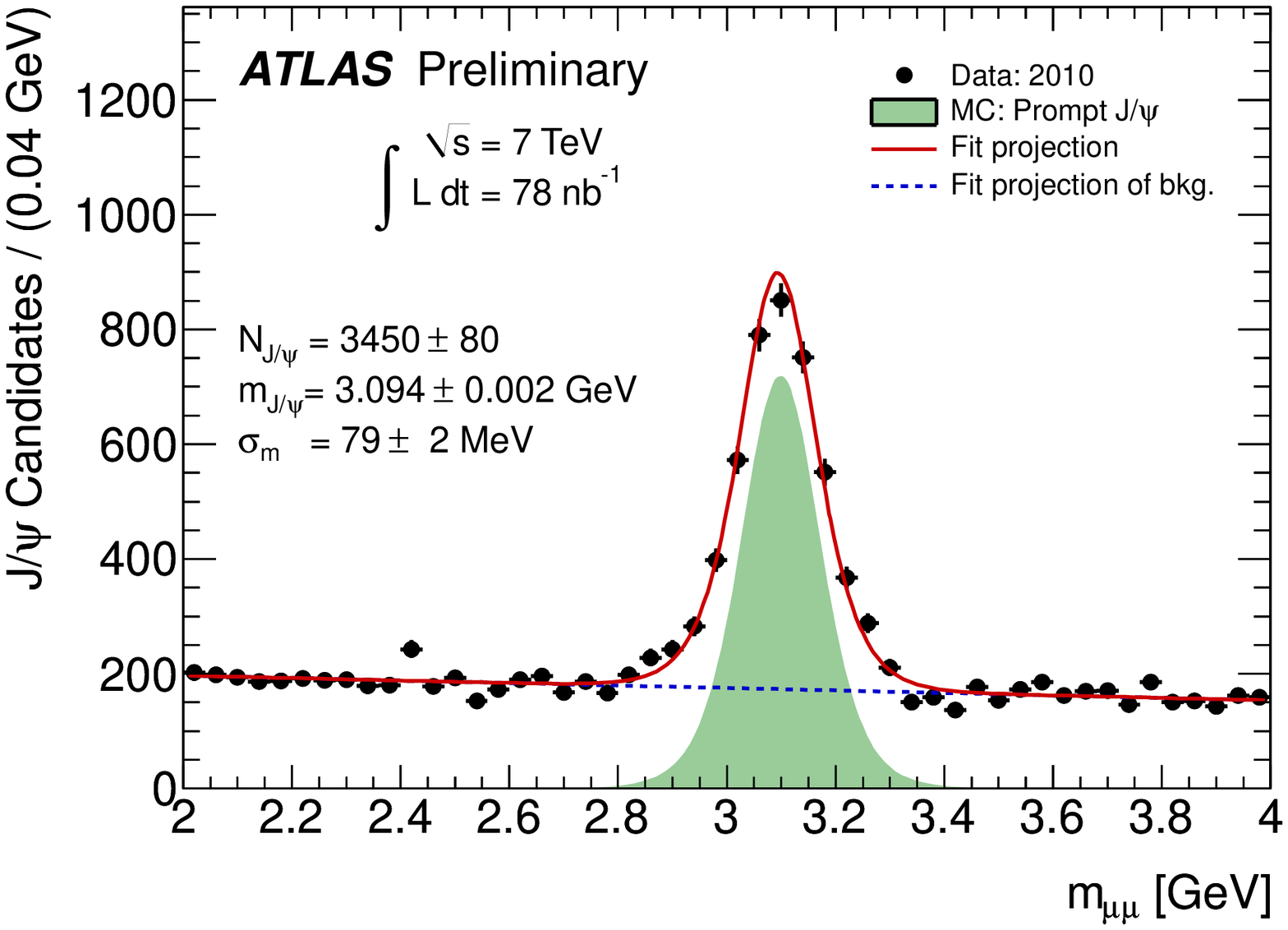}
\caption{ 
Invariant mass distributions of $J/\psi\rightarrow\mu^+\mu^-$ candidates in three muon $\eta$ categories: 
both muons in the barrel ($|\eta|<1.05$) (left), one muon in the barrel, the other in the end-cap ($1.05<|\eta|<2.5$) (middle), both muons in the end-cap (right). 
} \label{fig_05}
\end{figure*}

\begin{figure*}[t]
\centering
\includegraphics[width=77mm]{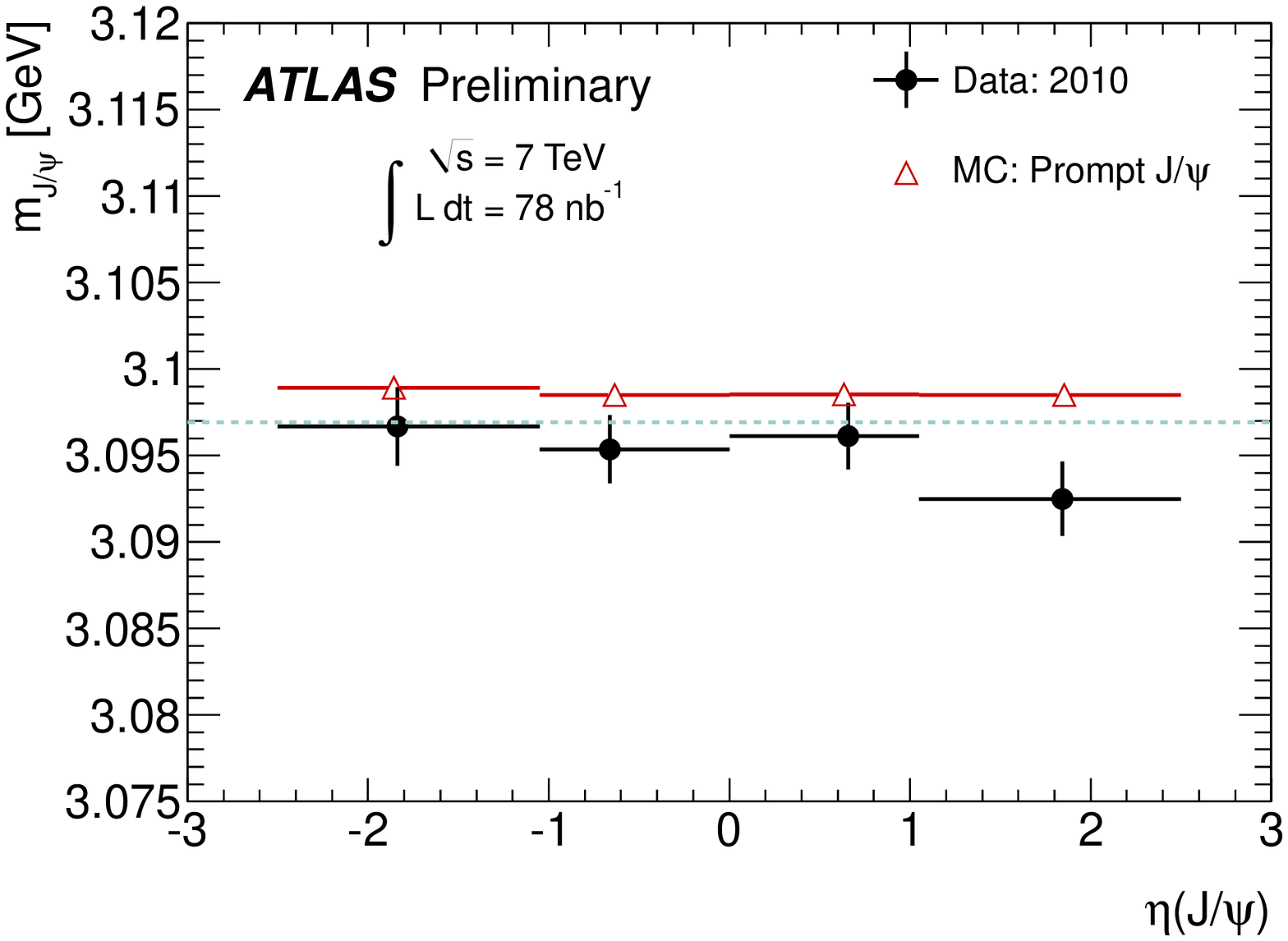}
\includegraphics[width=77mm]{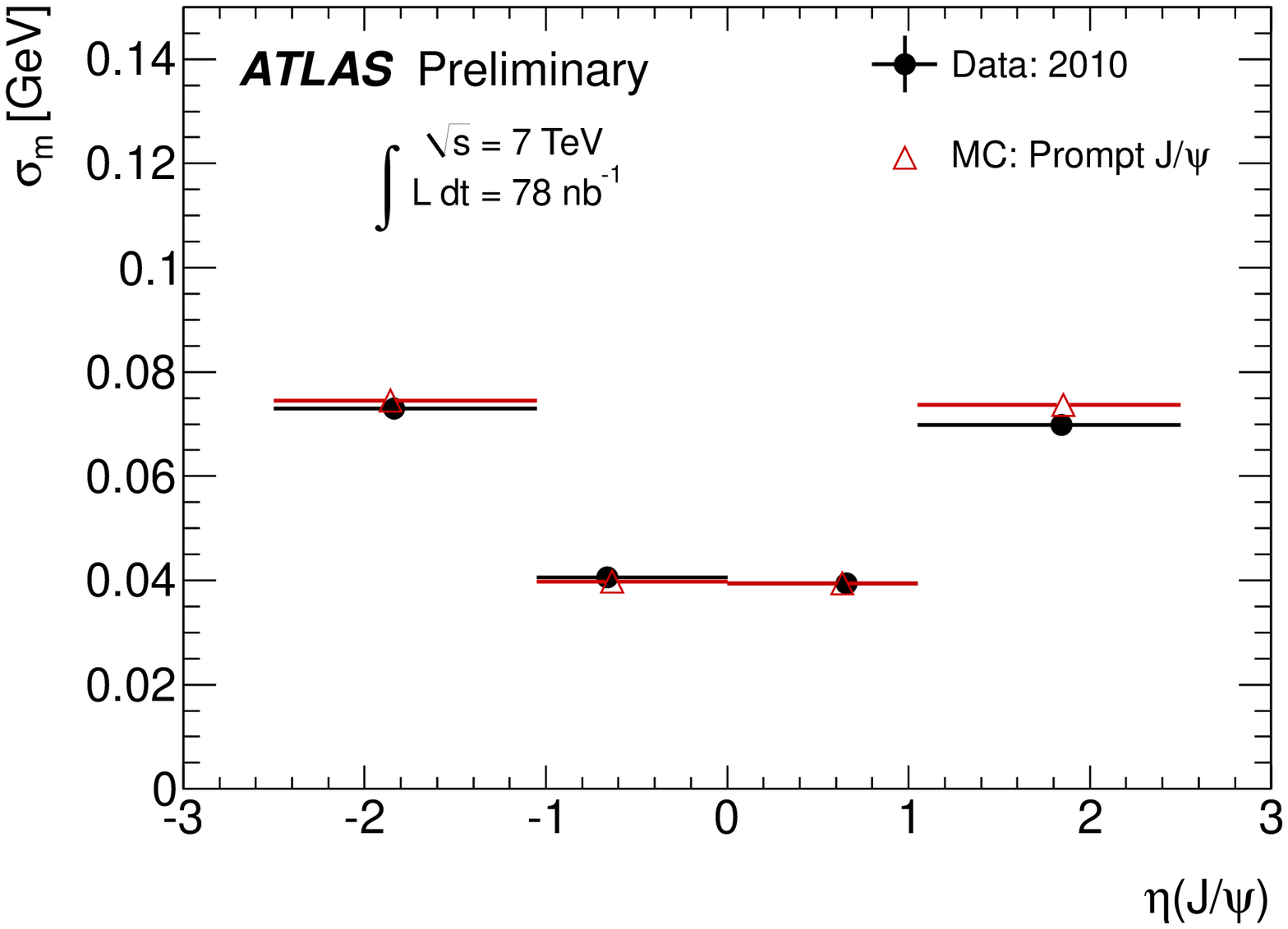}
\caption{
Reconstructed mass (left) and mass resolution (right) as a function of $\eta$ of $J/\psi$. 
The dashed line corresponds to the PDG value of the $J/\psi$ mass. 
} \label{fig_06}
\end{figure*}

The $J/\psi$ mass extracted from the di-muon invariant mass spectrum using ID tracks is used to study ID performance~\cite{ATLAS-CONF-2010-078}. 
To maximize the available statistics, an event is selected for further analysis if it passes an L1 muon trigger with no $p_T$ cut, or an HLT muon trigger with no $p_T$ cut seeded by an L1 MBTS trigger. 
Each event is required to contain at least one primary vertex built from at least three ID tracks, each of which containing at least one hit in Pixel and at least six in SCT. 
Only muons associated with ID tracks that have at least one hit in Pixel and six in SCT, and with a total momentum of at least $3~$GeV are accepted. 
The muon is labeled a barrel muon when its pseudo-rapidity, $\eta$, is $|\eta|<1.05$ and an end-cap muon when $1.05<|\eta|<2.5$. 
The two ID tracks from each pair passing the selections are fitted to a common vertex and a loose cut of $\chi^2<200$ on the the vertex fit is applied. 
Those pairs that successfully form a vertex are regarded as $J/\psi$ candidates. 
The candidates can be formed from muon pairs in which either muon can be tagged or combined. 
Figure~\ref{fig_05} shows the invariant mass distributions in three muon $\eta$ categories. 
An unbinned maximum likelihood fit with a Gaussian (signal) and a first order polynomial (background) is performed. 
The mass width with both muons in the end-cap region is about 2.5 times wider than one with both muons in the barrel. 
This behavior is caused by the higher amount of detector material and the shorter effective trajectory in the magnetic field in the forward region. 
Figure~\ref{fig_06} shows the mass (left) and the mass resolution (right) as a function of $\eta$ of $J/\psi$. 
The reconstructed masses in data and MC are well close to the PDG value. 
$J/\psi$ candidates that are observed in the forward regions have a wider mass resolution than those seen in the central region and this behavior is well described by MC. 

\subsection{Using $\Upsilon$ mesons} 
\begin{figure*}[ht]
\centering
\includegraphics[width=56mm]{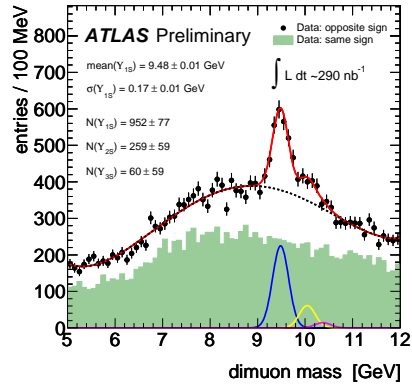}
\caption{
$\Upsilon(1S, 2S, 3S)\rightarrow\mu^+\mu^-$ candidates. 
Same sign pairs are shown as a control. 
} \label{fig_04}
\end{figure*}

Figure~\ref{fig_04} shows di-muon pairs with invariant masses between 5 and 12~GeV. 
Events which fire either of two triggers are admitted to the analysis; an L1 muon trigger with no $p_T$ cut, or an HLT muon trigger with a cut at $p_T$ of 4~GeV. 
The corresponding integrated luminosity is $290$~nb$^{-1}$. 
The applied offline $p_T$ cuts are $p_T>4$~GeV for leading muons and $p_T>2.5$~GeV for trailing muons. 
At least one muon in a pair is required to be a combined muon. 
An unbinned maximum likelihood fit with three Gaussians (signals) and fourth order polynomial (backgrounds) is performed. 
The Gaussians are with fixed same width and spacings fixed to the PDG values of three $\Upsilon$ states but the mass position is allowed to float in the fit. 
The fitted mass value of $\Upsilon(1S)$, $9.48 \pm 0.01$~GeV, is close to the PDG value ($9.4603 \pm 0.0003$~GeV). 

\section{$J/\Psi$ PHYSICS RESULTS}
Differential cross section of $J/\psi$ and the ratio of non-prompt to prompt $J/\psi$ in $pp$ collisions at $\sqrt{s}=7$~TeV are measured~\cite{ATLAS-CONF-2010-062}. 

\subsection{Differential cross section} 
Selections of $J/\psi$ is similar to those in subsection~\ref{withJpsi}. 
Furthermore, at least one muon in a pair is required to be a combined muon and only the HLT muon trigger with no $p_T$ cut seeded by an L1 MBTS trigger is used. 
Each event is weighted to correct acceptance ($A$), reconstruction efficiency ($\varepsilon_{reco}$) and trigger efficiency ($\varepsilon_{trig}$). 
\begin{equation}\label{eq:weight}
w^{-1} := A(p_T, y; \lambda_i) \times \varepsilon_{reco}(\vec{p}_{\mu^+}) \times \varepsilon_{reco}(\vec{p}_{\mu^-}) \times \varepsilon_{trig}(\vec{p}_{\mu^+}, \vec{p}_{\mu^-}) 
\end{equation}
Reconstruction efficiency is taken from MC. 
Trigger efficiency is measured with L1 MBTS triggered data. 
Acceptance depends on five independent variables, chosen as  $p_T$, $y$ and azimuthal angle $\phi$ of $J/\psi$ and two angles characterizing the $J/\psi$ polarization scenario ($\lambda_i$), $\theta^*$ and $\phi^*$. 
$\theta^*$ is the angle between the direction of the positive muon momentum in the $J/\psi$ decay frame and the momentum of $J/\psi$ in the $J/\psi$ frame. 
$\phi^*$ is the angle between the $J/\psi$ production and decay planes in the $J/\psi$ rest frame. 
The decay probability of $J/\psi$ is written as follows: 
\begin{equation}\label{eq:decay}
\frac{d^2N}{d\cos\theta^*d\phi^*} \propto 1+\lambda_\theta \cos^2\theta^* + \lambda_\phi \sin^2\theta^*\cos2\phi^*+\lambda_{\theta\phi}\sin2\theta^*\cos\phi^*. 
\end{equation}
Since the $J/\psi$ polarization has not yet been measured at the LHC, five extreme scenarios are used to span the theoretical uncertainty of the acceptance. 
The five scenarios are 
a flat polarization scenario ($\lambda_\theta=\lambda_\phi=\lambda_{\theta\phi}=0$), 
a longitudinal polarization scenario ($\lambda_\theta=-1$, $\lambda_\phi=\lambda_{\theta\phi}=0$) 
and three transverse polarization scenarios ($\lambda_\theta=+1$, $\lambda_\phi=-1, 0, +1$, $\lambda_{\theta\phi}=0$). 
The longitudinal (transverse) scenario has a larger (smaller) acceptance than the flat scenario. 
The flat scenario is adopted for the central value of the calculation. 
An unbinned likelihood fit of mass is performed in multiple bins of $p_T$ and $y$ of di-muon to determine the $J/\psi$ yield. 
A study of the various sources of systematics and their sizes can be found in~\cite{ATLAS-CONF-2010-062}.  

\begin{figure*}[t]
\centering
\includegraphics[width=77mm]{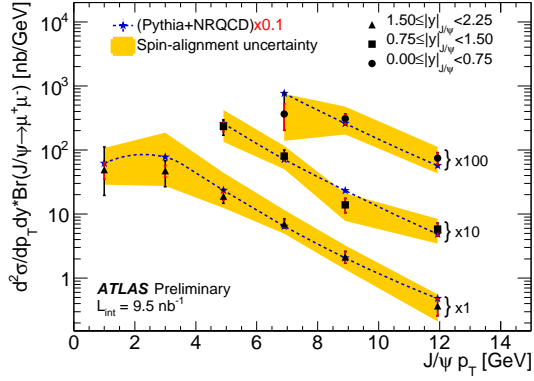}
\caption{
Corrected inclusive $J/\psi$ production cross section as function of $J/\psi$ $p_T$ and rapidity. 
Red bars are statistical errors and black bars are the quadrature sum of statistical errors and statistical errors. 
Overlaid is a band representing the variation of the result under various spin-alignment scenarios representing a theoretical uncertainty, 
and the prediction from PYTHIA MC using the ATLAS MC09 tune. 
The three $p_T$ distributions and corresponding PYTHIA predictions have been multiplied by the scaling factor indicated on the plot in order to visually separate the three sets of data for presentational reasons. 
Note also that for clarity the PYTHIA prediction in each divided by 10, so that it sits on the corresponding distribution rather than the one above. 
} \label{fig_07}
\end{figure*}

Differential cross section is shown in Fig.~\ref{fig_07}. 
Yellow bands represent the span of the extreme polarization scenarios. 
The results show good agreement in the $p_T$ and rapidity dependence of the PYTHIA prediction~\cite{PYTHIA, ATLASPYTHIATUNE}. 
The factor of 10 in absolute normalization with PYTHIA is being traced to tuning by ATLAS/the structure functions used. 

\subsection{Ratio of non-prompt to prompt $J/\psi$ production cross sections}
Pseudo-proper time ($\tau$) is used to resolve $J/\psi$ from $b$-decays. 
$L_{xy}$ is the displacement of a candidate from the primary vertex projected on the plane perpendicular to the beam axis. 
Using $L_{xy}$ and mass ($m$) and $p_T$ of $J/\psi$, $\tau$ is expressed as 
\begin{equation}\label{eq:pseudo-proper time}
\tau = L_{xy}m/p_T. 
\end{equation}
An unbinned event by event maximum likelihood fit of mass and $\tau$ in $p_T$ bins from $1$ to $15$~GeV is performed. 
Probability density functions describing $\tau$ distributions are $\delta$ function (prompt) plus exponential (non-prompt) convoluted with Gaussian (tracking resolution) for signal, 
and $\delta$ function plus two exponentials (first with positive slope, second symmetric for positive and negative $\tau$) convoluted with Gaussian. 
The background shape determined from side bands is used for the fit for the signal mass region within $\pm3\sigma$. 

\begin{figure*}[t]
\centering
\includegraphics[width=91mm]{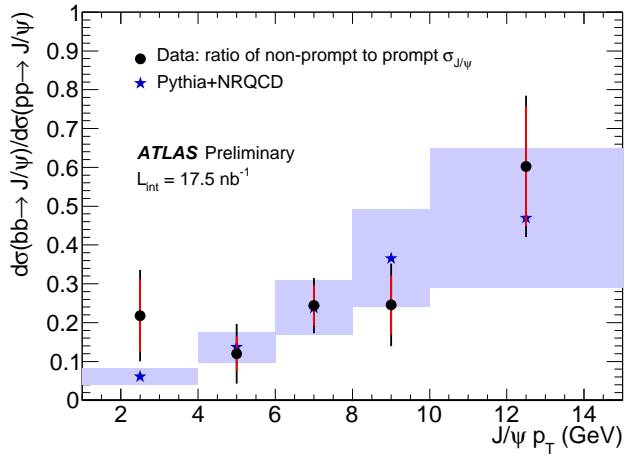}
\caption{
Ratio of non-prompt to prompt $J/\psi$ production cross sections as a function of $p_T$ of $J/\psi$. 
Overlaid is a band representing the prediction from PYTHIA MC. 
} \label{fig_08}
\end{figure*}

The ratio of non-prompt to prompt $J/\psi$ production cross sections is shown as a function of $p_T$ in Fig.~\ref{fig_08}. 
Data and PYTHIA MC are consistent within errors. 
On the experimental data points in Fig.~\ref{fig_08} the size of a total systematic error is shown, while the sources are discussed in~\cite{ATLAS-CONF-2010-062}.   

\section{SUMMARY} 
Quarkonia and $D$ mesons are reconstructed with the ATLAS detector with the expected mass resolutions. 
The differential cross section of $J/\psi$ in bins of $p_T$ and rapidity is measured. 
While the dependence on $p_T$ and rapidity of data agrees well that of PYTHIA, the normalization of PYTHIA is significantly larger than one of data. 

\begin{acknowledgments}
The author is supported in part by the Research Fellowship of the Japan Society for the Promotion of Science. 
\end{acknowledgments}

\end{document}